\documentclass[twocolumn,showpacs,superscriptaddress,aps]{revtex4-1}

\usepackage{amsfonts}
\usepackage{amsmath}
\usepackage{amssymb}
\usepackage{graphicx}
\usepackage{dcolumn}
\usepackage{dsfont}
\usepackage{times}
\usepackage{epsfig}
\usepackage[]{units}

\usepackage{lipsum} %% to make equation full width

\newcommand{\R}{{\mathbb R}}

\def\epsilon{\varepsilon}
    
\def\beq{\begin{equation}}
\def\eeq{\end{equation}}

\begin{document} \sloppy

\title{Traveling Waves in 2D Hexagonal Granular Crystal Lattices}
%\author{A. Leonard, C. Chong, P.G. Kevrekidis and C. Daraio}

%\author{        \and
%        C. Chong         \and
%        P.G. Kevrekidis         \and
%        C. Daraio      
%%}

\author{A. Leonard }
\affiliation{Department of Civil and Mechanical Engineering, California Institute of Technology, Pasadena CA 91125, USA}
\author{C. Chong}
\affiliation{Department of Mathematics and Statistics, University of Massachusetts, Amherst, MA 01003-4515, USA}
\author{P.G. Kevrekidis}
\affiliation{Department of Mathematics and Statistics, University of Massachusetts, Amherst, MA 01003-4515, USA}
\author{C. Daraio }
\affiliation{Department of Mechanical and Process Engineering (D-MAVT), 8092 Zurich, Switzerland}

%\institute{A. Leonard \at
%	Department of Civil and Mechanical Engineering, California Institute of Technology, Pasadena CA 91125, USA\\
%           C. Chong \at
%           Department of Mathematics and Statistics, University of Massachusetts, Amherst, MA 01003-4515, USA\\
%            \and            
%           P.G. Kevrekidis \at
%              Department of Mathematics and Statistics, University of Massachusetts, Amherst, MA 01003-4515, USA\\
%               \and   
%           C. Daraio \at
%              Department of Mechanical and Process Engineering (D-MAVT), 8092 Zurich, Switzerland\\
%}

%\date{\today}


\begin{abstract}
We describe the dynamic response of a two-dimensional hexagonal packing of uncompressed stainless
steel spheres excited by localized impulsive loadings. After the initial impact strikes the system, a characteristic wave structure emerges and 
continuously decays as it propagates through the lattice. Using an extension of the binary collision approximation (BCA) for one-dimensional chains, we predict its decay rate, which compares well with numerical simulations and experimental data. While the hexagonal lattice does not support constant speed traveling waves, we provide scaling relations that characterize the power law decay of the wave velocity. Lastly, we discuss the effects of weak disorder on the directional amplitude decay rates.
\end{abstract}

\maketitle
%============================================================================
\section{Introduction}\label{sec:intro}

Granular crystals can be defined as ordered closely packed arrays of elastically interacting particles.  The fundamental wave propagation in one-dimensional (1D) homogeneous granular chains has been a topic of numerous investigations over the past few years, see e.g. the reviews~\cite{Nesterenko2001,Sen08,PGK11} and references therein. In 1D, granular chains support the formation of highly localized traveling waves, which can be described by relatively simple numerical and analytical approaches ~\cite{Nesterenko2001,ahnert09,pego05}. The highly nonlinear response of these crystals stems from the Hertzian contact interaction between compressed spheres, and the absence of tensile response between the grains \cite{Johnson:1987}. In this setting, the presence or absence of static precompression has been used to control the intrinsic wave propagation properties of impulsive stresses \cite{Nesterenko2001,Daraio2006bb}.  In addition to their prototypical Fermi-Pasta-Ulam type functional form and features (such as the existence of supersonic traveling waves), granular crystals have been proposed for different applications including shock and energy absorbing layers~\cite{Daraio2006bb,Hong2005,Fraternali2008,Doney2006}, actuating devices \cite{dev08}, acoustic lenses \cite{Spadoni}, acoustic diodes \cite{Nature11} and sound scramblers \cite{dar05,Nesterenko2005}.

% \begin{figure*}[t]
%\begin{center}
%\includegraphics[scale=0.45]{Fig2a.eps} 
%\includegraphics[scale=0.45]{Fig2b.eps} 
%%\epsfig{file=Fig2a.eps,width=.45 \textwidth}
%%\epsfig{file=Fig2b.eps,width=.45 \textwidth} 
% \end{center}
% \caption{(a) Velocity profile for the first ten beads in an initially at rest 1D granular crystal after
% impact on the left boundary with $A=d=M=1$.  Notice that after the
% first three beads, the maximum velocity attained by each bead is approximately constant after impact (denoting the
% rapid settling of the chain into a traveling solitary wave). 
% (b) Velocity profile for the first ten beads along the zero degree line of an initially at rest 2D hexagonal granular crystal with $A=d=M=1$. 
% Notice that the maximum velocity attained by each bead decays (denoting
% the spreading of the energy to an increasing number of beads). 
% }
%  \label{fig:decay_profiles}
%  \end{figure*}

Although one-dimensional granular chains have been studied at considerable length, homogeneous higher dimensional crystals have only recently started to be  explored in more detail. Examples of the topics examined include: the dynamic load transfer path in disordered two-dimensional systems~\cite{l6,l7,Owens:2011,Kondic:2012}, the stress-wave anisotropy in centered-square nonlinear chains with different materials~\cite{l10,l11,Aswasthi}, and the redistribution of energy in a square lattice with one or more interstitials~\cite{ivan12}. Only a few studies have investigated the dynamic response of basic two-dimensional (2D) hexagonal particle arrays \cite{l8,Abd:2010,l9,Nishida}, focusing primarily on the weakly nonlinear (i.e. compressed) system response \cite{Coste:2008,Gilles:2003}. Although the theme of excitation propagation is of fundamental interest in its own right, it may also be of significant practical interest. For instance, in applications such as sound absorption, the goal is often to  find conditions that promote energy dispersion (or dissipation) and it is thus of interest to examine such properties of different types of  bead arrangements in higher dimensions. In that light, we are interested in the dynamic response of impacting an initially at rest hexagonal lattice.

In one spatial dimension, traveling solitary waves are generated upon striking one end of an initially at
rest chain of particles \cite{Nesterenko2001,Lindenberg}. The properties of these solitary waves depend on the composition of the chain, on the geometry and material properties of the particles and on the degree of static precompression applied to the chain. The two main analytical  approximations
developed in order to study  such traveling waves are based on a quasi-continuum  approximation \cite{Nesterenko2001,ahnert09}, where a partial differential equation is derived  from the pertinent lattice model, and a binary collision approximation (BCA). The BCA relies on the assumption that at a given time, a significant portion of the energy of the traveling structure is centered over two lattice sites where the resulting equations can then be solved exactly \cite{Lindenberg}. 
We should note in passing that in addition to these methods and especially
in 1D, the Fourier space, co-traveling frame analysis of~\cite{pego05} enables
a computation of the exact traveling wave up to a prescribed numerical 
tolerance.
In 2D configurations, one can study traveling structures by observing a row of beads along different radial directions from the impact bead. For a square packing, it has been shown that quasi-one-dimensional traveling solitary waves emerge upon striking the lattice \cite{Leonard11}, and are essentially described by the one-dimensional theory. In other arrangements, such as hexagonal packings, we argue that such quasi-one-dimensional motion is impossible, since each adjacent row and column will be affected upon being struck, regardless of the striking angle. After initially impacting a single bead, the energy will gradually be spread over progressively more and more lattice sites. Since the energy continues to spread to an increasing number of beads, the (energy and) velocity magnitude will gradually decrease, and thus a perfect traveling solitary wave will be impossible to support. This is in contrast to the 1D (resp. 2D square) situation, where the amplitude of the velocity profile remains almost constant, and hence, supports a traveling solitary wave. Thus, hexagonally packed granular crystals are far better candidates for applications such as sound absorption.

In this paper, we use numerical, theoretical, and experimental tools  in order to study how energy is spread throughout a hexagonally packed granular crystal lattice upon being struck at different ``strike angles'' and being observed at different ``observation angles''. In addition to confirming the absence of a traveling solitary wave, we identify scaling power law patterns for the decay of the velocity as the wave travels through multiple layers of beads within the structure. In a simple special case, we are able to generalize the BCA into a ternary collision approximation (TCA), which yields a reasonable numerical (and approximate semi-analytical) estimate for the velocity decay. Lastly, additional numerical simulations incorporating weak disorder in the particle lattice are performed in order to more realistically simulate and compare with the experimental results.

\begin{figure}
 \centerline{
\includegraphics[scale=0.55]{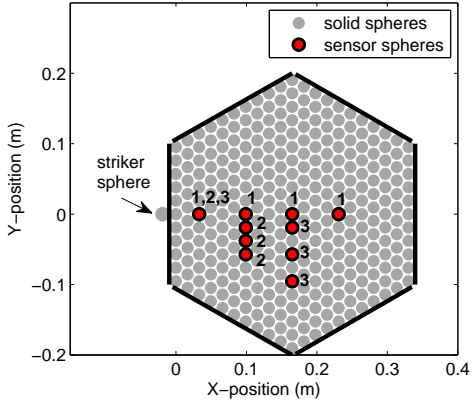} 
}
 \caption{Schematic diagram of the experimental setup. The solid stainless steel sphere locations are represented with grey circles. The confining walls are represented by solid black lines. The striker sphere impacts the system as shown, with an initial velocity $V_{\rm striker}$ in the $x$-direction. The locations of sensor particles are marked with a red circle and outlined in black. The number next to each sensor location indicates the different sensor configurations: 1, 2, and 3 (with 4 sensors present for any given experiment).}
 \label{fig:expsetup}
 \end{figure}

%==========================================================================
\section{Experimental Setup}\label{sec:exp}

We performed experiments on a 2D hexagonal array of particles consisting of 11 spheres along each edge of the lattice (see Fig.~\ref{fig:expsetup}). The spheres were stainless steel (type 316), with diameter $d=19.05~mm$ \cite{mcmaster}. For subsequent calculations and numerical simulations, the particles were assumed to have a Young's modulus $E=193~GPa$, Poisson's ratio $\nu=0.3$, and density $\rho=8000~\nicefrac[]{kg}{m^{3}}$  \cite{Leonard11}. The hexagonal particle lattice was confined, but not compressed, by walls on all 6 sides, with a hole at the impact location along one edge. Since we are concerned with only the primary traveling excitation, a comparatively ``soft'' material, delrin, was chosen to line the confining walls ($E=3.1~GPa$, $\nu=0.35$, and $\rho=1400~\nicefrac[]{kg}{m^{3}}$  \cite{Leonard11}) in order to delay reflections from the system edges. To aid in the lattice particle alignment, a slight tilt ($<5^{0}$) was imposed in the x-direction of the experimental setup.

To excite the system, a striker sphere, identical to the spheres composing the 2D hexagonal granular crystal, impacted the central particle along one edge. The striker sphere rolled through a channel down the inclined experimental setup and its initial velocity was calculated from the drop height. To measure the motion of individual particles within the system we used custom fabricated sensor particles that consisted of a miniature triaxial accelerometer (PCB 356A01 with sensitivity 0.51 $\nicefrac[]{mV}{\frac{m}{s^{2}}}$) embedded in a stainless steel sphere. A detailed description of the sensor particles can be found in \cite{Leonard11}. Based on the data acquisition system used (NI BNC-2110 and NI PCI-6123 with simultaneous sampling rate at $500~\nicefrac[]{kS}{s}$), the number of sensors present in the array during a single experimental impact was limited to four (with data acquisition in both the  $x$- and $y$-directions). Therefore, to better capture the system response, three different sensor configurations were used: (1) along the $0^{0}$ observation direction, (2) perpendicular to and near the impact, and (3) perpendicular to and further from the impact (see Fig. \ref{fig:expsetup}). The sensor in the lattice location closest to the impact was present in all three sensor configurations.

Previous studies on the effects of weak disorder in tightly packed 2D granular arrays showed that the exact system response depends on the initial contact lattice, and is generally consistent between repeated impacts on an individual configuration \cite{l11}. In order to capture the variability caused by differences in initial contact lattices, 20 different particle packings were assembled and experimentally tested (for each sensor configuration, 1-3) with an initial striker velocity of $V_{\rm striker}=0.40~\nicefrac[]{m}{s}$. Each of the 20 initial assemblies was impacted repeatedly to calculate average wave front amplitude and arrival time values, which could differ due to slight variations in impact conditions, such as exact alignment and speed of the striker particle, and minor particle rearrangements. Additionally, experiments for 20 different initial particle packings with sensor configuration 1 were performed for striker velocities, $V_{\rm striker}=0.25~\nicefrac[]{m}{s}$ and $V_{\rm striker}=0.70~\nicefrac[]{m}{s}$.

\begin{figure*}[t]
 \centerline{
%  \centerline{\epsfig{file=thelattice.eps,width=.45 \textwidth} 
%\epsfig{file=thelattice_TCA.eps,width=.45 \textwidth} 
\epsfig{file=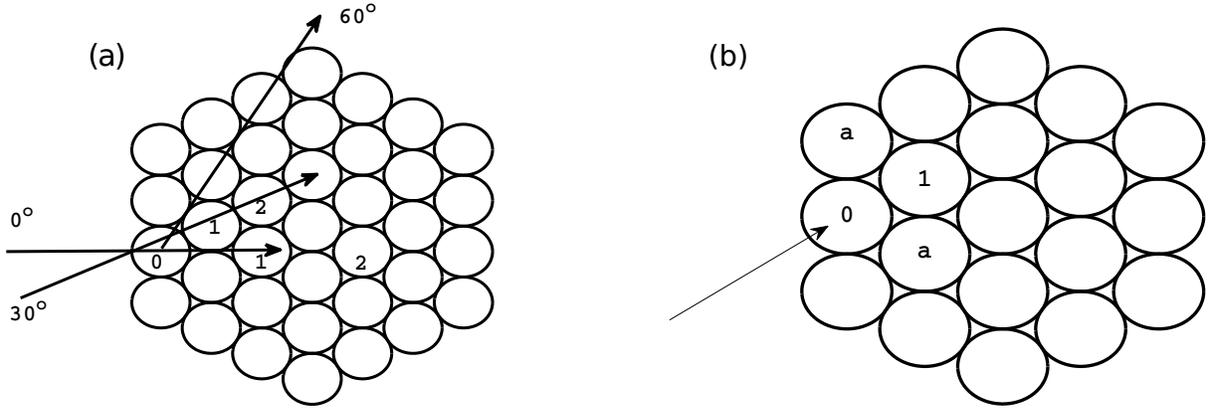,width=.9 \textwidth} 
}
 \caption{
(a) Orientation of the hexagonal lattice and possible striking and observation angles. The numeric labels correspond
to the counting conventions used along the respective observation angles. (b) Labeling convention for the ternary collision approximation}
 \label{fig:beads}
 \end{figure*}

The impact conditions of the experimental assembly were chosen for experimental ease and consistency. In the numerical simulations and theoretical considerations below, the initial velocity $V_{0}$ was assigned to the $n=0$ bead, that is the bead impacted by the striker sphere in experiments (see also Fig. \ref{fig:beads}). Numerical simulations, comparing the system response for an array with $V_{\rm striker}=V_{0}$ and $V_{n=0}=V_{0}$, showed that the difference was negligible for the studied system.

%==========================================================================
\section{Model} \label{sec:model}

In order to model the above experimental setup, we consider an ordered homogeneous lattice of uncompressed spherical beads arranged in a hexagonal configuration, neglecting, at least for the purposes of considerations herein, 
dissipative effects. The Hamiltonian of such a system has the form,

\begin{equation}
\begin{split}
 H = \sum_{m,n}  \frac{1}{2} M p_{m,n}^2 + V( | d e_1 + q_{m+2,n} - q_{m,n}|) \\+V( | d  e_2 + q_{m+1,n+1} - q_{m,n}|)\\ + V( | d e_3 + q_{m+1,n-1} - q_{m,n}|)
 \end{split}
\end{equation}
where $q_{m,n} \in \R^2$ is the displacement of the $\{m,n\}$ bead from its initial position, $d$ is the bead diameter, $M$ is the bead mass, $e_1 = [  2,0  ]^T, e_2 = [ 1, \sqrt{3} ]^T, e_3 = [ 1, -\sqrt{3}  ]^T $ and
\begin{equation}
V(r) = A \frac{2}{5} [ d - r ]^{5/2}_{+} 
\end{equation}
where $A$ is a parameter depending on the elastic
properties of the material and the geometric characteristics of the 
beads~\cite{Nesterenko2001}, and the bracket is defined by $[r]_+ = \max(0,r)$.
All parameters can be rescaled to unity using the transformation,
%$$\tilde{q}_{m,n} = d\,q_{m,n}  \quad \quad \tilde{t} = \alpha t \quad \quad \alpha = d^{1/4} \sqrt{A/M}.$$ \label{scaling}
\begin{equation}
 \tilde{q}_{m,n} = d\,q_{m,n},  \quad \quad \tilde{t} = \sqrt{\frac{M}{A \,d^{1/2}}} \, t  \label{eq:scaling}
 %\tilde{q}_{m,n} = d\,q_{m,n},  \quad \quad \tilde{t} = d^{-1/4} \sqrt{M/A} \, t  \label{eq:scaling}
\end{equation}
where the variables with the tilde represent the solutions in the physical scaling (i.e., the dimensional ones). 
We define $q_{m,n} = [ x_{m,n} , y_{m,n} ]^T$ and $\dot{q}_{m,n} = p_{m,n} = [ u_{m,n}, v_{n,m}]^T$.
In these variables, the equations of motion have the form,
\begin{widetext}
\begin{equation} \label{eq:model}
\left .
\begin{array}{ll} \displaystyle
\dot{x}_{m,n} =& u_{m,n} \\ 
\dot{y}_{m,n} =& v_{m,n} \\
\dot{u}_{m,n} =&  \frac{V'( r_1) \, r^{x}_1}{r_1} + \frac{V'( r_2) \, r^{x}_2}{r_2} - \frac{V'( r_3) \, r^{x}_3}{r_3} 
                - \frac{V'( r_4) \, r^{x}_4}{r_4} - \frac{V'( r_5) \, r^{x}_5}{r_5} + \frac{V'( r_6) \, r^{x}_6}{r_6} \\
\dot{v}_{m,n} =&  \frac{V'( r_1) \, r^{y}_1}{r_1} + \frac{V'( r_2) \, r^{y}_2}{r_2} - \frac{V'( r_3) \, r^{y}_3}{r_3} 
                - \frac{V'( r_4) \, r^{y}_4}{r_4} - \frac{V'( r_5) \, r^{y}_5}{r_5} + \frac{V'( r_6) \, r^{y}_6}{r_6} \\
                 \end{array}
                 \right\}
\end{equation}
{where $ r_j = \sqrt{  (r_j^x)^2 + (r_j^y)^2 }$ for $j=1 \dots 6$ and}
%
%\begin{align*}
%&r_1^x =  \cos(0) + x_{m+2,n} - x_{m,n}    \qquad\qquad  r_1^y     = \sin(0) + y_{m+2,n} - y_{m,n} \\
%&r_2^x =  \cos(\pi/3) + x_{m+1,n+1} - x_{m,n}     \qquad  r_2^y    =  \sin(\pi/3)  + y_{m+1,n+1} - y_{m,n} \\
%&r_3^x =  \cos(-\pi/3) - x_{m-1,n+1} + x_{m,n}     \quad  r_3^y    =  \sin(-\pi/3)  - y_{m-1,n+1} + y_{m,n} \\
%&r_4^x =  \cos(0) - x_{m-2,n} + x_{m,n}     \qquad\qquad  r_4^y =      \sin(0) - y_{m-2,n} + y_{m,n} \\
%&r_5^x =  \cos(\pi/3) - x_{m-1,n-1} + x_{m,n}     \qquad  r_5^y    =  \sin(\pi/3) - y_{m-1,n-1} + y_{m,n}\\ 
%&r_6^x =  \cos(-\pi/3) + x_{m+1,n-1} - x_{m,n}     \quad  r_6^y    =  \sin(-\pi/3) + y_{m+1,n-1} - y_{m,n} 
%\end{align*}
%\begin{center}

%\begin{widetext}
$$
\begin{array}{ll}
r_1^x =  \cos(0) + x_{m+2,n} - x_{m,n}           & r_1^y  = \sin(0) + y_{m+2,n} - y_{m,n} \\
r_2^x =  \cos(\pi/3) + x_{m+1,n+1} - x_{m,n}     & r_2^y  =  \sin(\pi/3)  + y_{m+1,n+1} - y_{m,n} \\
r_3^x =  \cos(-\pi/3) - x_{m-1,n+1} + x_{m,n}    &  r_3^y =  \sin(-\pi/3)  - y_{m-1,n+1} + y_{m,n} \\
r_4^x =  \cos(0) - x_{m-2,n} + x_{m,n}           &  r_4^y =      \sin(0) - y_{m-2,n} + y_{m,n} \\
r_5^x =  \cos(\pi/3) - x_{m-1,n-1} + x_{m,n}     &  r_5^y =  \sin(\pi/3) - y_{m-1,n-1} + y_{m,n}\\ 
r_6^x =  \cos(-\pi/3) + x_{m+1,n-1} - x_{m,n}    &  r_6^y =  \sin(-\pi/3) + y_{m+1,n-1} - y_{m,n} 
\end{array}
$$
\end{widetext}
%\end{center}

 \begin{figure*}[t]
  \centerline{
\epsfig{file=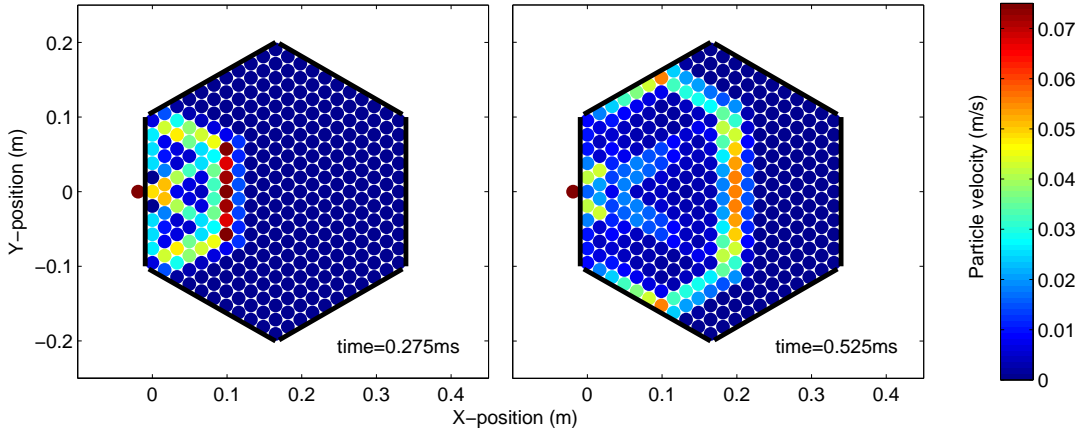, width=.9 \textwidth} 
 }
 \caption{Numerical simulation results showing the wave front shape evolution for the 11-sphere edge system tested in experiments. The array was centrally impacted along the edge by a striker sphere with initial velocity $V_{\rm striker}=0.40~\nicefrac[]{m}{s}$. The colorbar indicates particle velocity magnitude in $\nicefrac[]{m}{s}$.}
 \label{fig:smallsim}
 \end{figure*}

\begin{figure*}
  \centerline{
\epsfig{file=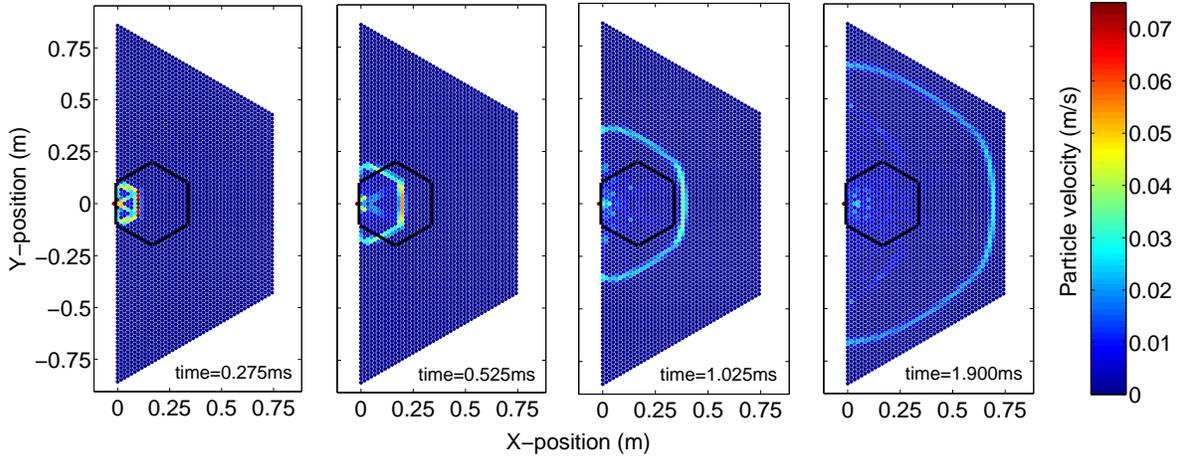, width= .9 \textwidth} 
 }
 \caption{Numerical simulation results showing the wave front shape evolution for a larger hexagonal packing (91 particles along the left, long edge, and 45 particles along the shorter edges). The array was centrally impacted along the edge by a striker sphere with initial velocity $V_{\rm striker}=0.40~\nicefrac[]{m}{s}$. The black hexagon indicates the wall locations for the size of the experimental setup. The colorbar indicates particle velocity magnitude in $\nicefrac[]{m}{s}$.}
 \label{fig:largesim}
 \end{figure*}

We can solve this system numerically using a standard 4th order Runge-Kutta method. The delrin-lined walls present in experiments were modeled as spheres with an infinite radius. The numerical simulation results for the configuration corresponding to the experimental setup are shown in Fig.~\ref{fig:smallsim}. Additionally, numerical simulations were performed for a larger array consisting of 45-spheres along the bounding hexagonal edge, compared to the 11-sphere per edge system tested in experiments. 
Figure~\ref{fig:largesim} shows the evolution of the wavefront shape over the larger 2D hexagonal system. 
It is interesting to observe therein, in addition to the continuous decrease
of the wavefront speed, its gradual deformation in what appears to be 
a semi-circular shape. We will return to this point later in the text
(in the discussion after presenting our detailed results).
Although this model ignores dissipation and disorder in the system, it has been found to provide a good comparison with the experimental observations, as we also detail below.

%---------------------------------------------------------------------------------------------------------------------------------
\section{Ternary Collision Approximation: A Theoretical Approach for $30^\circ$ Angle of Observation} \label{sec:TCA}

System~\eqref{eq:model} is extremely complex. Indeed, many theoretical studies of two-dimensional lattice equations are for simpler, single component
systems~\cite{Wattis06,Wattis07,Wattis07b,Quan}, although traveling waves
in fully 2D Hamiltonian square lattices of springs have also been examined; see e.g.~\cite{frieseckem}. 
To help reduce the complexity of the system we consider a setting which is amenable to a semi-analytical description. More specifically, we turn to an analog of the BCA that has been developed for 1D chains \cite{Rosas2003,Rosas2004,Rosas2004b,Harbola,Lindenberg}. We first assume that the beads travel approximately along fixed angles upon impact, based on symmetry considerations. From a theoretical point of view, it is more straightforward to develop the theory for a striking and observation angle of $30^\circ$ since the pattern does not alternate along this line of observation, in contrast to what is the case for a $0^\circ$ striking, see Fig.~\ref{fig:beads}. Considering the time scale where only the beads adjacent to the struck bead are affected results in a so-called ternary collision approximation (TCA). Actually, there are four beads involved in the distribution of the energy at each collisional step, yet we use the symmetry to reduce the number of degrees of freedom to three. In the renormalized system, we have the following TCA for the system
of the right panel of Fig.~\ref{fig:beads},
\begin{eqnarray*}
 \ddot{x}_0 &=& -[ x_0 - x_1 ]_+^{3/2} -  2 \cos(\theta) [ \cos(\theta) x_0 - x_a ]_+^{3/2} \\ 
 \ddot{x}_1 &=&  [ x_0 - x_1 ]_+^{3/2} \\
 \ddot{x}_a &=&  [ \cos(\theta) x_0 - x_a ]_+^{3/2} 
\end{eqnarray*}
where  $x_0$ is the displacement along the impact direction, $x_1$ is the displacement along that same direction of
the bead adjacent to the impacted bead, and $x_a$ is the displacement of the bead adjacent to the impacted bead 
at the angle  $\theta$.  By symmetry, the bead adjacent along the $-\theta$ line will have the same contribution as the $x_a$ bead.
Now define $z = x_0 - x_1$ and $y= \cos(\theta) x_0 - x_a$. We have then
\begin{equation} \label{eq:TCA_strain}
\left. 
\begin{array}{ll}
\ddot{z} =& -2  [z]_+^{3/2} -  [y]_+^{3/2} \\ 
 \ddot{y} =&  - [ z ]_+^{3/2}/2 -  3 [y]_+^{3/2} /2  
\end{array} \right\},
\end{equation}
where we used $\theta = \pi/3$, which corresponds to a hexagonal packing.  Although System~\eqref{eq:TCA_strain} is remarkably simpler than the
original System~\eqref{eq:model}, it only lends itself to an exact
(analytically obtainable) solution in the case where
$y(t)=\omega z(t)$, where $\omega\in\R$. However, note that this prescription requires also that
the initial conditions satisfy the same scaling. The definition of
$z$ and $y$ necessitate (from the corresponding initial 
conditions) that $\omega=1/2$. However, using the above ansatz in System~(\ref{eq:TCA_strain}) yields $\ddot{z}=-(2+\omega^{3/2}) [z]_+^{3/2}$, as well as
the ``compatibility condition'' $2+\omega^{3/2}= (1/2+3/2 \omega^{3/2})/\omega$, whose sole
real solution is $\omega \approx 0.382$. From the above, it is clear that
the compatible solution is close to (although not exactly) satisfying
the initial conditions. In that light, we will also use for
comparison the exact analytical solution of the form:
\begin{eqnarray}
 t= _2F_1([2/5,1/2],[7/5],\frac{2 \sqrt{2+{\omega^{3/2}}} z^{5/2}}{5 E}) \frac{z}{\sqrt{2 E}},
\label{hyperg}
\end{eqnarray}
where $E$ in the relevant hypergeometric function expression 
stands for the constant ``energy'' of the oscillation associated
with $z(t)$ and can be computed as $E=v^2/2$, where $v$ is the 
initial velocity of the bead $x_0$.

As explained in \cite{Lindenberg}, the main purpose of the BCA is to offer an 
approximation of the velocity of the traveling wave  
that propagates through the chain. In order to compute this ``pulse velocity''
one needs to define the time $T_n$ the pulse resides on bead $n$.
Let $t=0$ be the time when the velocities of beads $n-1$ and $n$ become equal.
Then $T_n$ is the time when the velocities of beads $n$ and $n+1$ become equal and
the pulse velocity observed at the $nth$ bead is $c_n= 1/T_n$.

In order to approximate the residence time $T_n$, the BCA combines two points of view.
First, it is assumed that the interaction is not instantaneous, such that the velocity
of the impacted bead will gradually decrease. However, the BCA is only valid if two
beads are involved in the dynamics, which is clearly not the case over the entire
residence time $T_n$. Thus, strictly in terms of the BCA, it is not clear at all
how to initialize the next step in the iteration. To circumvent this issue,
it is supposed that the interaction is instantaneous such that the conservation
of kinetic energy and momentum can be used to compute the velocity that bead $n$ will 
emerge with after the collision with bead $n-1$. For example, for a heterogeneous 1D chain,
these conservations yield  $V_n = 2 V_{n-1}/(1 + m_{n}/m_{n-1})$, where $m_n$ is the
mass of bead $n$ \cite{Lindenberg}. Applying
this relationship recursively yields
\begin{equation} \label{eq:BCA_decay}
V_n = V_0 \prod_{j=1}^{n}  \frac{2}{1 + m_{j}/m_{j-1}} \, .
\end{equation}
Thus, the reduced equations in the BCA setting at the $nth$ step are initialized with
$z(0) = 0, \dot{z}(0) = V_n$. Although it appears somewhat awkward to adopt these two viewpoints in order to make the approximation
work, the results in 1D chains have been surprisingly good~\cite{Rosas2003,Rosas2004,Rosas2004b,Harbola,Lindenberg}. 
For this reason we carry out a similar procedure to compute the pulse velocity of the traveling wave in the hexagonal lattice.

Due to the complexity of the hexagonal system there is no
clear way to obtain an analog of relationship~\eqref{eq:BCA_decay} using conservation of kinetic energy and momentum. However,
in both the 1D BCA and 2D TCA setting, we can compute
the time it takes for the initially impacted bead to obtain a velocity equal
to its adjacent partner. For a homogeneous 1D chain, the BCA predicts this to occur for a velocity equal
to $50\%$ of the initial velocity. Thus, to understand what contribution the additional beads in the hexagonal configuration will absorb, we use the reduced equations~\eqref{eq:TCA_strain} to compute when the velocities of beads 0 and 1 are equal (see Fig.~\ref{fig:beads} for the labeling convention). The ratio
of the velocity at this time compared to the 1D case is one way to quantify the fraction of momentum that is absorbed by the additional beads. Call this value $c$. For example, suppose the velocity at the time of intersection was predicted to be $0.45$ in the TCA and
$0.5$ in the BCA. Then $c=0.9$, since the bead only reached $90\%$ of the value it would have in the absence of the additional beads. 
If in addition, we assume that this ratio remains constant as the traveling structure propagates through the lattice, we expect that bead $n$ emerges
with the velocity $V_n =  c V_{n-1}$ after collision with bead $n-1$. Applying this relationship recursively yields,   
\begin{equation} \label{eq:TCA_decay}
V_n =   V_{0} c^n,  
\end{equation}
where $V_0$ is the impact velocity. Equation~\eqref{eq:TCA_decay} will now play the role of~\eqref{eq:BCA_decay} for the hexagonal system. In the rescaled system
we found $c\approx0.8452$.

\begin{figure*}
  \centerline{
\epsfig{file=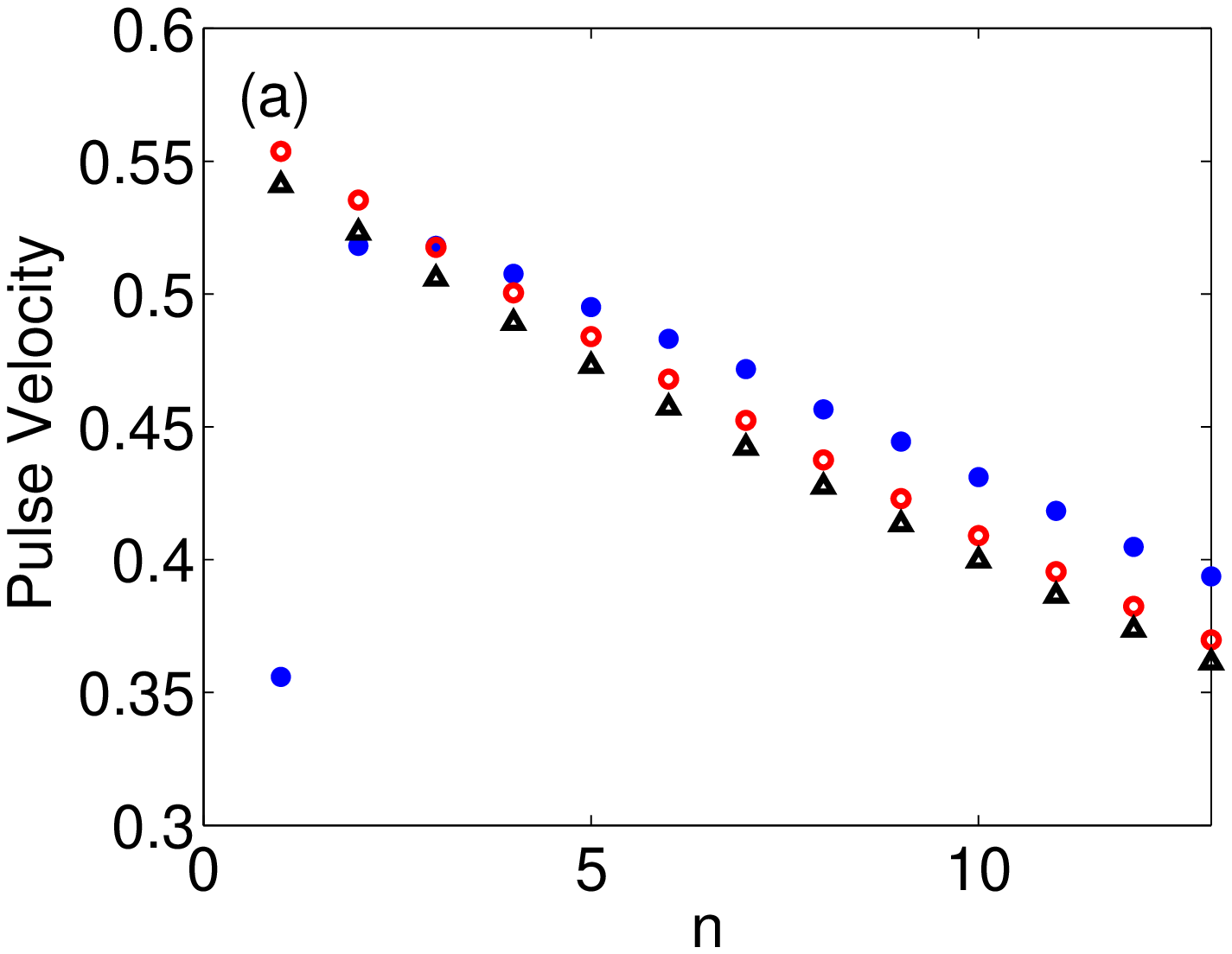,width=.4 \textwidth} 
\epsfig{file=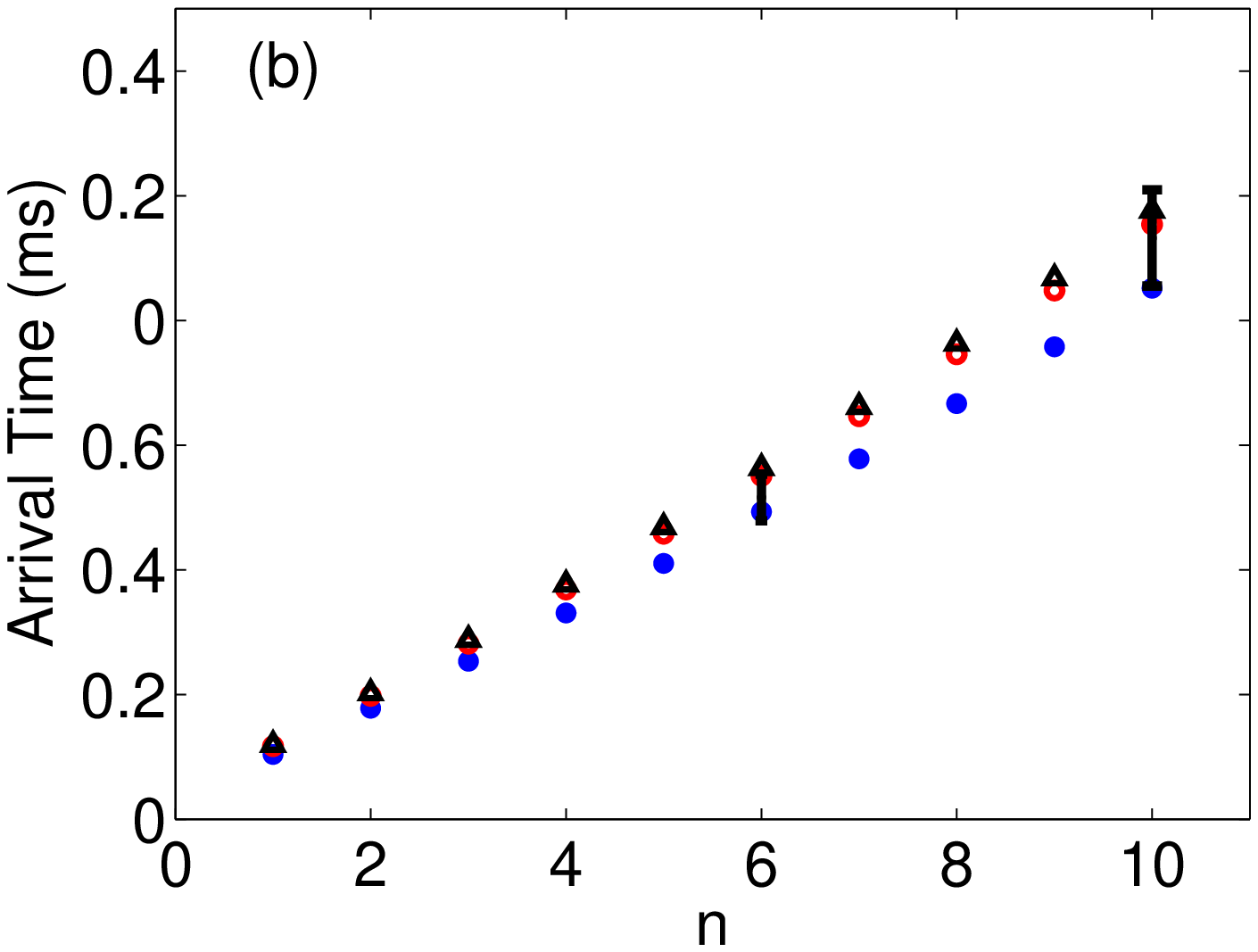,width=.4 \textwidth} 
}
 \caption{(a) Observed pulse velocity of the 
propagating structure in the ideal parameter case at various bead numbers $n$ for a $30^\circ$ strike and $30^\circ$ observation angle. The pulse velocity is defined as $c_n = 1/T_n$ where $T_n$ is the $nth$ residence time. The blue dots represent the calculated time based on numerical simulations and the red circles are the prediction based on the TCA. The black triangles denote the exact analytical solution to the TCA system (which,
however, is only approximate). The initial strike velocity was $V_0 = 0.1$. 
(b) The arrival time of the pulse for the experimental parameter values at a $0^\circ$ strike and $30^\circ$ observation angle with $V_0 = 0.4 ~\nicefrac[]{m}{s}$. The blue dots represent numerical simulations, the red circles are the TCA, the black triangles denote the exact analytical solution to the TCA, and the points with error bars at $n=6$ and $n=10$ are
the experimental values. 
   }
 \label{fig:TCA}
 \end{figure*}

Given the repetition of the fundamental building block 
of the right panel of Fig.~\ref{fig:beads}, we now
simply solve the TCA System~\eqref{eq:TCA_strain} with initial values defined through~\eqref{eq:TCA_decay} and compute the intersection time.
Doing this for several bead locations yields a
discrete set of approximations for the 
pulse velocity observed at those bead locations. See the left panel of Fig.~\ref{fig:TCA}
for an example with $V_0 = 0.1$. There, it can be seen that such an 
approximation of this average pulse velocity favorably compares 
with the full numerical results for the hexagonal lattice.

To amend the TCA to a $0^\circ$ strike (but still observing along the $30^\circ$ line)
we need to understand how the energy is transferred among the first two beads that are in contact
with the impact bead (see Fig.~\ref{fig:beads}). Due to symmetry considerations, it is reasonable to conjecture that the
velocity contribution along the $30^\circ$ (resp. -$30^\circ$) will be half of the impact velocity. 
This would be the case if kinetic energy and momentum were conserved. This relation was observed to be fairly accurate (see the right panel of Fig.~\ref{fig:vodecay0}).
At the $0^\circ$ strike and $30^\circ$ observation angle we have experimental data available for comparison, see the right panel of Fig.~\ref{fig:TCA}. 
Due to the limited number of sensors available, we simply
present the arrival time data (which is the sum of the relevant residence times). We should also note that the experiments
use the arrival time at the first sensor position in Fig.~\ref{fig:expsetup} to define $t=0$. Thus, in order to compare the experimental values
to the TCA and numerical simulations we estimate the arrival time at the first sensor using the numerical simulation.

%---------------------------------------------------------------------------------------------------------------------------------
\section{A Numerical \& Experimental Study: Velocity Decay Rate for Variable Striking/Observation Angle Combinations} \label{sec:numstudy}

 \begin{figure*}
\centerline{\epsfig{file=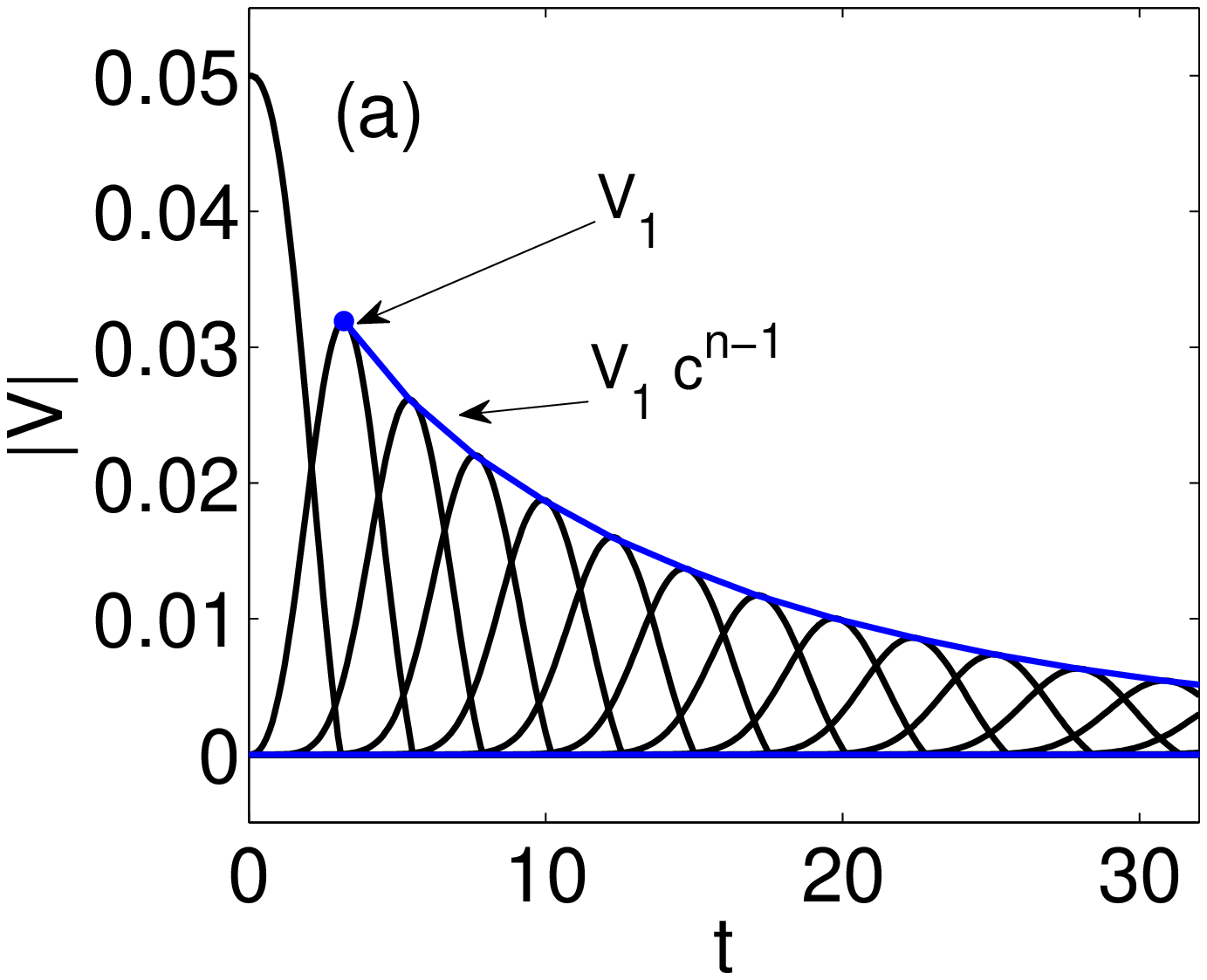,width=.4 \textwidth, height= .2 \textheight} 
\epsfig{file=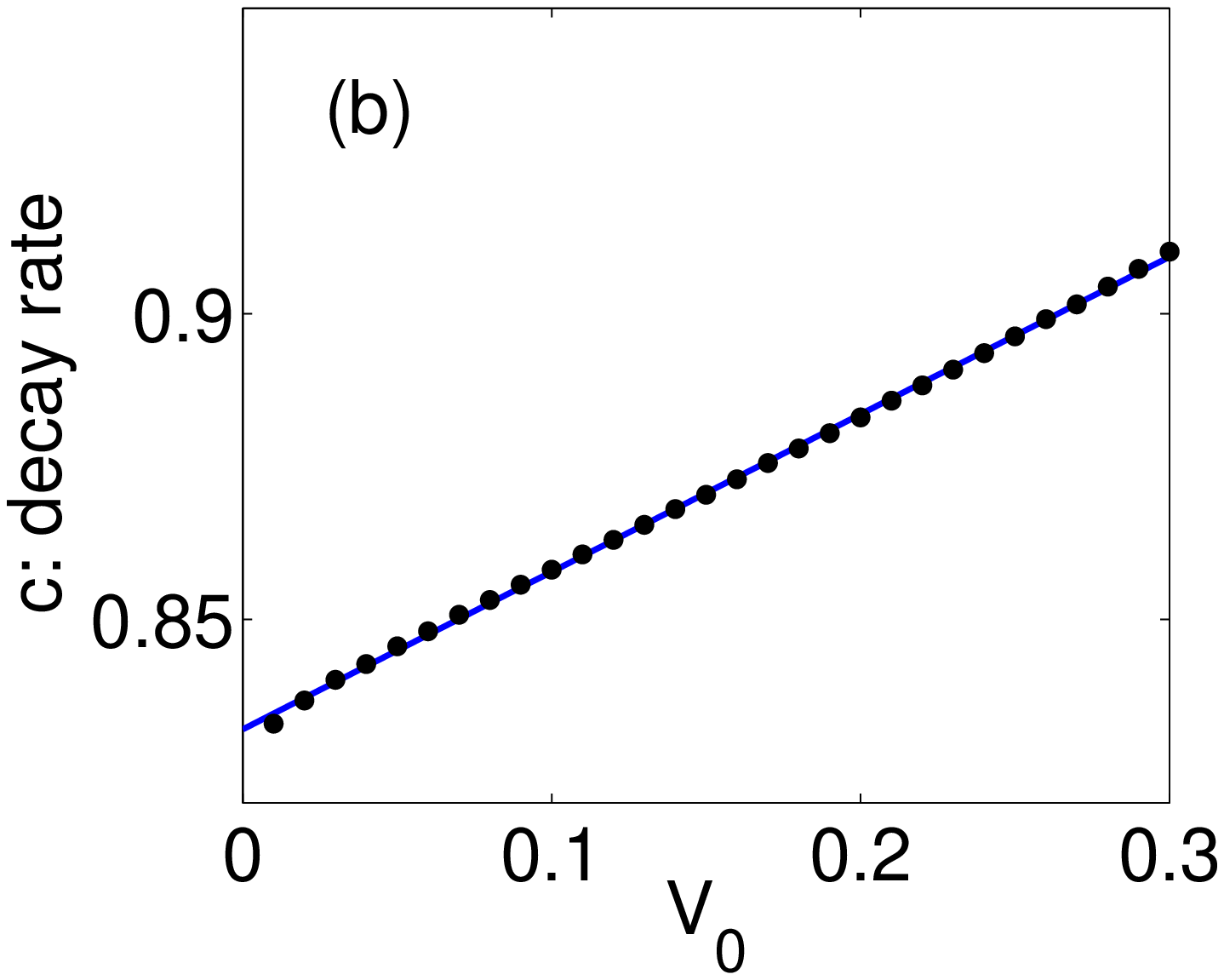,width=.4 \textwidth, height= .2 \textheight} }
\centerline{
\epsfig{file=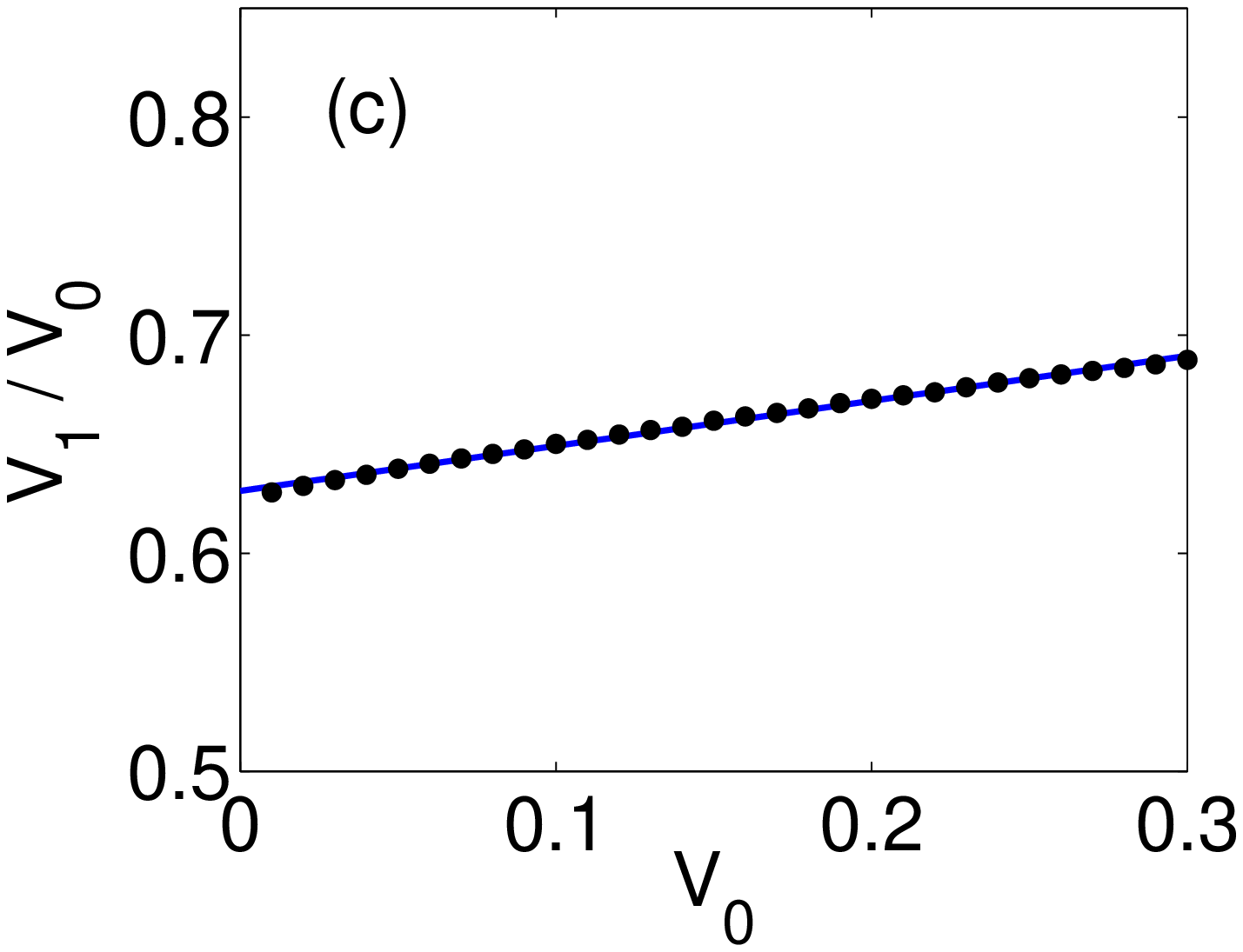,width=.4 \textwidth, height= .2 \textheight}
\epsfig{file=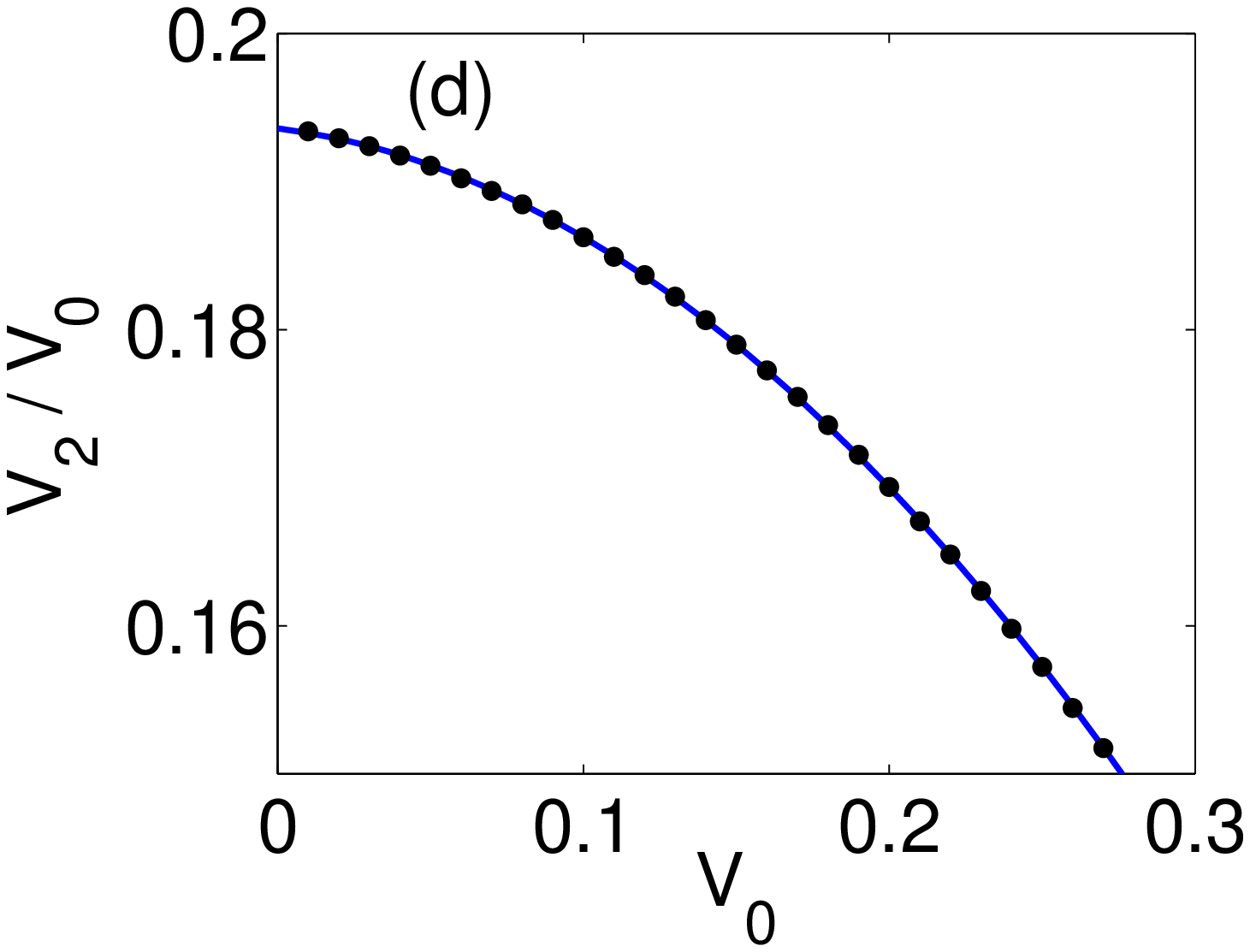,width=.4 \textwidth, height= .2 \textheight}
}

\caption{Illustration of the procedure to obtain the scaling law for the decay of the velocity profile. 
This example is for strike and observation on the $30^\circ$ line. (a)  Magnitude of velocity of beads $0-12$
versus time in the ideal parameter case with $V_0=0.05$. The maxima are fit with a function of the form
$ V_1 c^{n-1}$, where $V_1$ is the maximum velocity of the first bead (labeled in the plot). This
processes is repeated for several values of $V_0$ to obtain a sequence of decay rates $c(V_0)$.
(b) The sequence of decay rates $c(V_0)$ are fairly accurately fit with a linear function (see the text). (c)
The relationship between $V_0$ and $V_1$ is computed, and can be either linear (like in this case)
or quadratic. Finally, putting these relationships together yields a formula for the maximum velocity
at a given bead, see e.g. Eq.~\eqref{eq:v30}. (d) For a $0^\circ$ observation, the fitting begins at bead 2, where
the relationship between the initial striking velocity $V_0$ and $V_2$ is quadratic.
}
 \label{fig:vodecay30}
 \end{figure*}

In applications, not only is the time of arrival of the traveling wave important, but
the magnitude of its amplitude as well. The TCA, although simple enough to afford
analytical approximations, ignores the (weak) dependence of the decay rate $c$ on $V_0$ 
and is restricted to an observation angle of $30^\circ$. We wish to have a more accurate and robust description of the velocity decay rate
and thus we now turn to a (strictly) numerical study of other combinations of striking/observation angles.
We corroborate the obtained qualitative and quantitative
results by means of experimental observations.

\subsection{$30^\circ$ Strike:} 
In panel (a) of Fig.~\ref{fig:vodecay30}, the magnitude of the velocity of each 
bead along the $30^\circ$  line is shown against time. Notice how the maximum 
velocity obtained for each bead decreases as we move further from the impact point, as 
expected. We fit these maximum points with a function of the form $ V_1 c^{n-1}$ where
$c$ is a decay parameter that will now depend on the initial strike velocity $V_0$
(in contrast to the situation above which had c independent of $V_0$).
We start the fit at $V_1$ since the initial impact will only have
a velocity component, and thus the dynamics of the collision of the $n=0$ and $n=1$
bead are inherently different than the dynamics of the $n$ and $n+1$ beads for $n>0$.
In the latter case, the bead will have a velocity and position component.
In panel (b) of Fig.~\ref{fig:vodecay30} we fit the decay rates $c=c(V_0)$
with a linear function  $m V_0 + b$ with two free parameters $m,b$.
We found that $m\approx0.257$ and $b\approx 0.832$. In panel (c) of Fig.~\ref{fig:vodecay30} 
we obtain the relationship between $V_0$ and $V_1$, which has the form 
$V_1 = \alpha V_0^2 + \beta V_0$. We found $\alpha \approx 0.206$ and $\beta \approx 0.629$,
such that the decay between $V_0$ and $V_1$ is greater than that between
$V_n$ and $V_{n+1}$ for $n>0$.
We have then that,
\begin{equation} \label{eq:v30}
 V_n = V_0 \,( \alpha V_0 + \beta ) ( m V_0 + b )^{n-1}, \quad \quad n>0
\end{equation}
where $V_n$ is maximum velocity of the bead lying $n$ beads away from the strike point on the $30^{\circ}$ line. Clearly this description of the velocity
decay is more accurate than~\eqref{eq:TCA_decay}  since the 
dependence on the initial impact velocity is taken into account.
Nevertheless, we still observe that a power law decay of the
maximum velocity of each bead provides an accurate description of
the dynamics along the $30^\circ$ observation angle in the hexagonal chain.

\begin{figure*}
  \centerline{
\epsfig{file=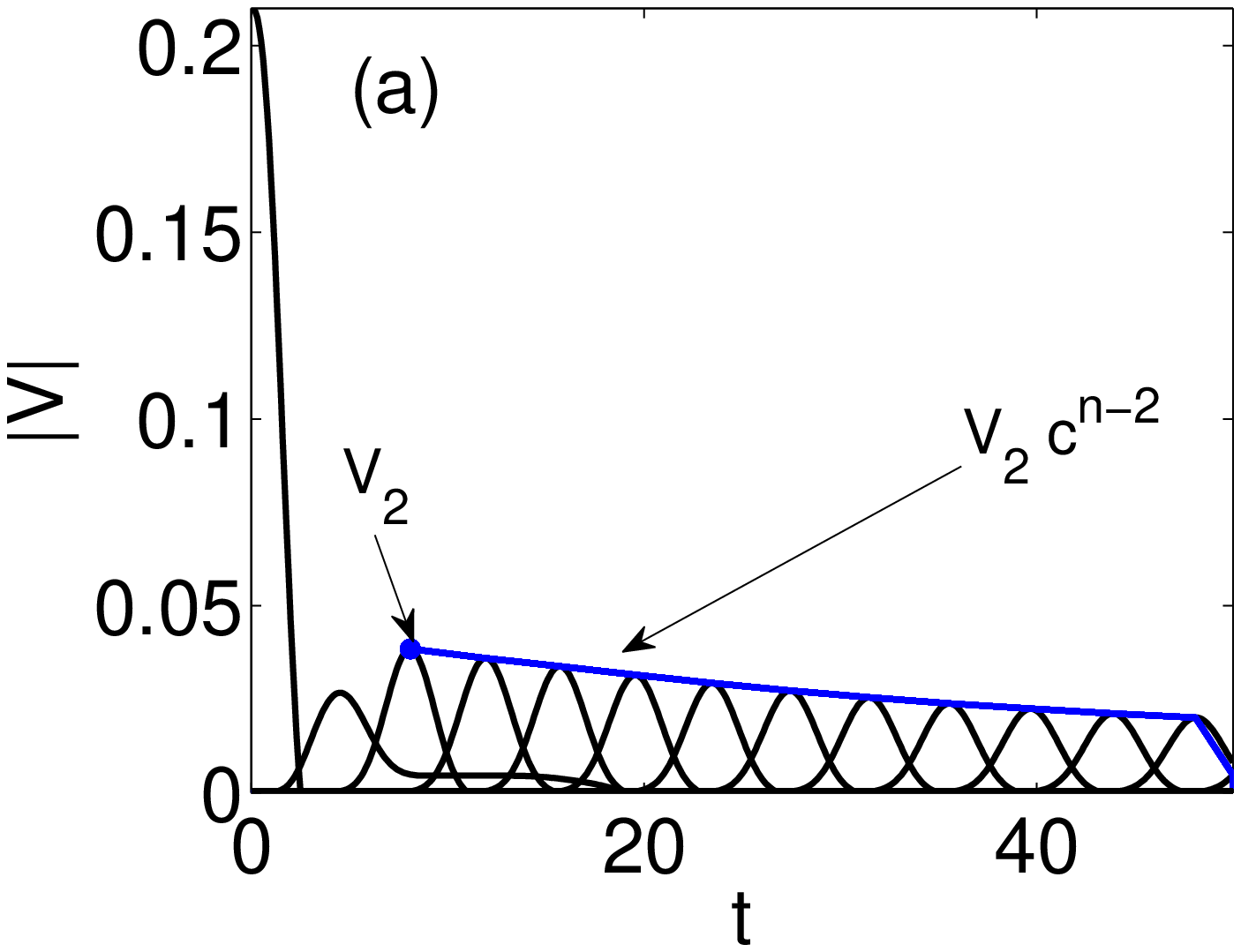,width=.4 \textwidth, height=.2 \textheight} \hspace{.2cm}
\epsfig{file=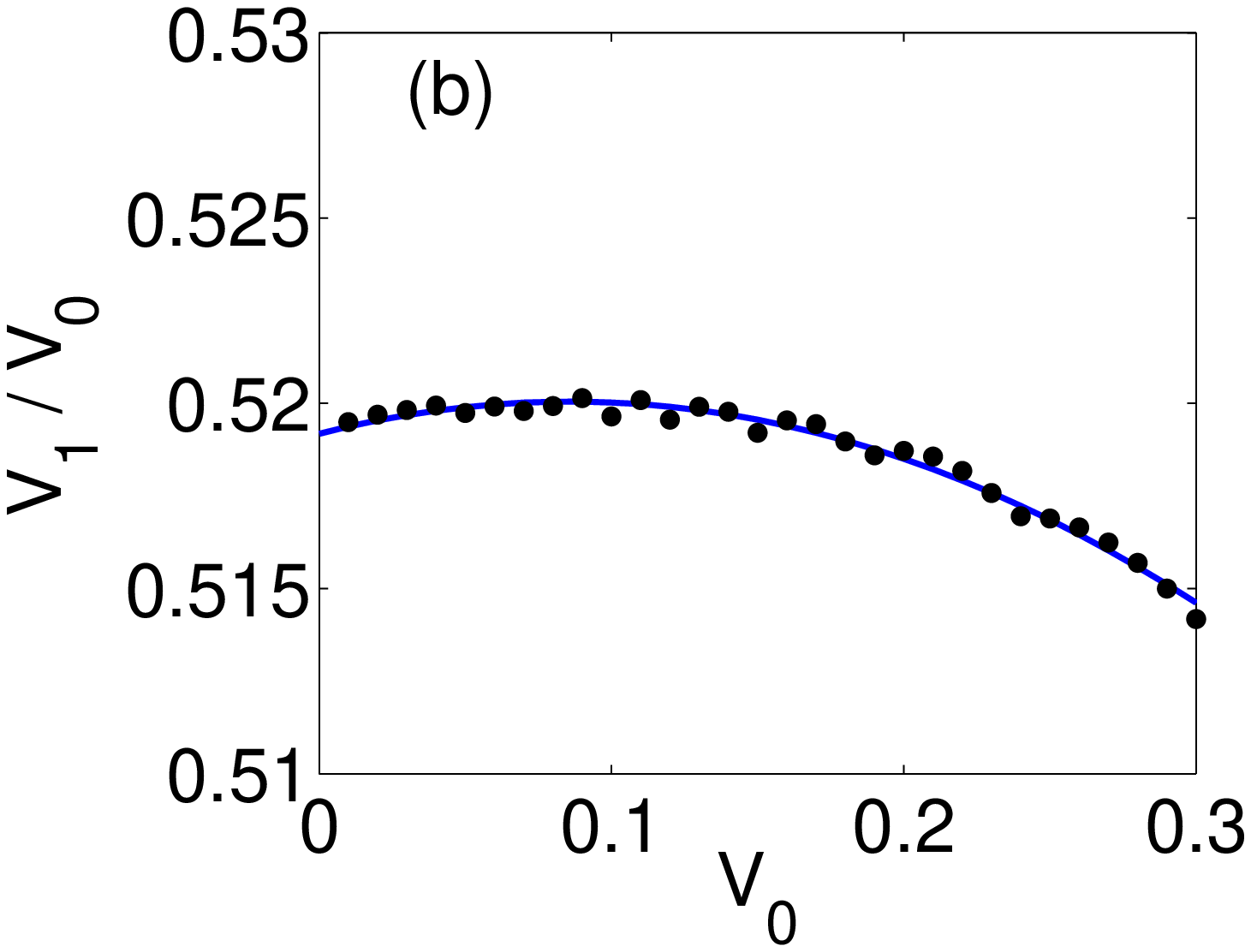,width=.4 \textwidth, height=.21 \textheight}   %internal note: Fig8c changed to Fig8b and Fig8b to Fig7d
 }
 \caption{(a) magnitude of the velocity of beads $0-12$ along the $0^\circ$ line upon being
struck at a $0^\circ$ angle. In this case, the fitting is more accurate starting at the second peak $V_2$. 
(b) The relationship between the initial striking velocity $V_0$ and $V_1$ for a
$0^\circ$ strike and $30^\circ$ observation is quadratic. For this particular case, we see that $V_1 \approx V_0/2$.}
 \label{fig:vodecay0}
 \end{figure*}

When observing along the zero degree line it turns out that the fitting is more accurate when starting
at bead 2, see Fig.~\ref{fig:beads} for the labeling convention. Panel (a) of Fig.~\ref{fig:vodecay0} shows a relavent example
where the velocity fitting starts at bead 2, but for a $0^\circ$ strike (which is analyzed in Sec.~\ref{30_0}). Since the relationship between $V_1$ and $V_0$ was linear, it is
natural to suppose that the relationship between $V_2$ and $V_0$ is quadratic, i.e.
$V_2 \approx  \alpha V_0^2 + \beta V_0 + \gamma $ (see panel (d) of Fig.~\ref{fig:vodecay30}
for example).  We have then that,
\begin{equation} \label{eq:v0}
V_n = V_0 \,( \alpha V_0^2 + \beta V_0 + \gamma ) ( m V_0 + b )^{n-2}, \quad \quad n>1 
\end{equation}
where $V_n$ is the magnitude of the maximum velocity of the $nth$ bead from 
the strike point on the $0^{\circ}$ line. Our fitting yielded $m\approx .3039$, $b\approx.8756$ and 
$\alpha \approx -0.4788, \beta \approx   -0.0256, \gamma \approx  0.1936$.

\subsection{$0^\circ$ Strike:}  \label{30_0}
Finally, we carry out the procedure for the case of striking at a $0^\circ$ angle. As mentioned above for a zero degree observation, the first bead behaves 
somewhat
differently (see panel (a) of Fig.~\ref{fig:vodecay0}), and therefore we start the fitting again at bead 2. 
We found that our scaling relations continue to be valid for a $0^\circ$ strike, but now with
$m \approx 0.1766$, $b \approx 0.8878$ and $\alpha \approx -0.2643,  \beta \approx  -.0969,  \gamma \approx .2152$.
The optimal scaling along the $0^\circ$ observation angle has the same form
as~\eqref{eq:v0}.

When observing at a $30^\circ$ angle,
the formula that yielded the best fit had the form,
\begin{equation} \label{eq:veq0}
V_n = V_0 \,( \alpha V_0^2 + \beta V_0 + \gamma ) ( m V_0 + b )^{n-1}, \quad \quad n>0 
\end{equation}
with $m \approx 0.1484 $,   $b \approx 0.8393$ and  $\alpha \approx -0.1182,  \beta \approx  0.0203 ,  \gamma \approx   0.5192$.
However, for the range of initial velocities used, the relationship between $V_1$ and $V_0$ is approximately
$V_1 = V_0/2$ (see panel (b) of Fig.~\ref{fig:vodecay0}) which yields the simplified formula,
\begin{equation} 
V_n = \frac{V_0}{2}( m V_0 + b )^{n-1}, \quad \quad n>0. 
\end{equation}

Note, in order to obtain the scaling law for arbitrary parameter values, one simply replaces each
occurrence of $V$ with $V  \sqrt{M/A}\,/d^{5/4} $.

%---------------------------------------------------------------------------------------------------------------------------------
\section{Numerical \& Experimental Study Incorporating Weak Disorder}

\subsection{Preparation Scheme}

Numerical simulations incorporating weak disorder were also performed for the 2D hexagonal system described in Section \ref{sec:exp}. Similar to \cite{l11}, weak disorder was incorporated in the simulations by assigning each particle in the array a random diameter, based on a normal distribution with mean $\mu=d=19.05~mm$ and standard deviation $\sigma=6tol$, where $tol=0.0127~mm$ is the manufacturer specified diameter tolerance of the particles used in experiments. The initial resting positions of the particles in the weakly disordered lattice were found in two steps. First, the particles were given an initial hexagonal lattice spacing assuming $d=d+tol$, to avoid large repulsive forces between overlapping particles. To bring the particles in contact, a $5N$ force was applied to each of the edge particles and an artificial damping (of the type introduced in \cite{DEV}) was used to settle the random particle motion. Secondly, the wall positions were found based on the slightly compressed settled configuration.
Then, gravity was introduced along the x-direction (in agreement with the experimental tilt), the 5N force was removed, and the particles were again settled through a similar, but weaker damping process. Once the initial positions were obtained, the settled array was impacted with a striker sphere, $V_{\rm striker}=0.25,~0.40,$ or $0.70~\nicefrac[]{m}{s}$, along one edge of the system. This process was performed for 20 different initial lattice configurations. For the subsequent discussion, we will use the terms ``ideal''  and ``weakly disordered'' to distinguish between numerical simulations where the spheres are all assigned the same diameter,  and those just described with a variable diameter.

%-------------------------------------------------------------------------------------------------------------------------------
\subsection{Discussion of Results}

\begin{figure*}
  \centerline{
\epsfig{file=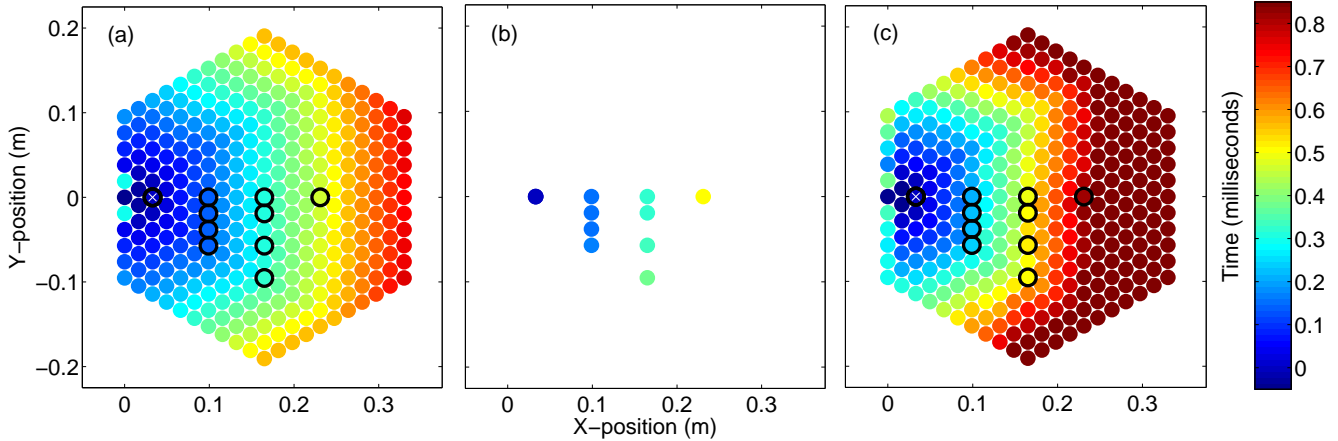, width=1.05 \textwidth} 
 }
 \caption{Wave front shape based on the relative wave front arrival time with respect to the particle marked with a white x. (a) Numerical simulation of ideal system (without disorder or dissipation). (b) Average wave front arrival time calculated from experiments. (c) Average wave front arrival time calculated from simulations with weak disorder. Black circles in the left and right panels indicate experimental sensor locations. The colorbar indicates the relative wave front arrival time in milliseconds.}
 \label{fig:expnumsim}
 \end{figure*}

\begin{figure*}
  \centerline{
\epsfig{file=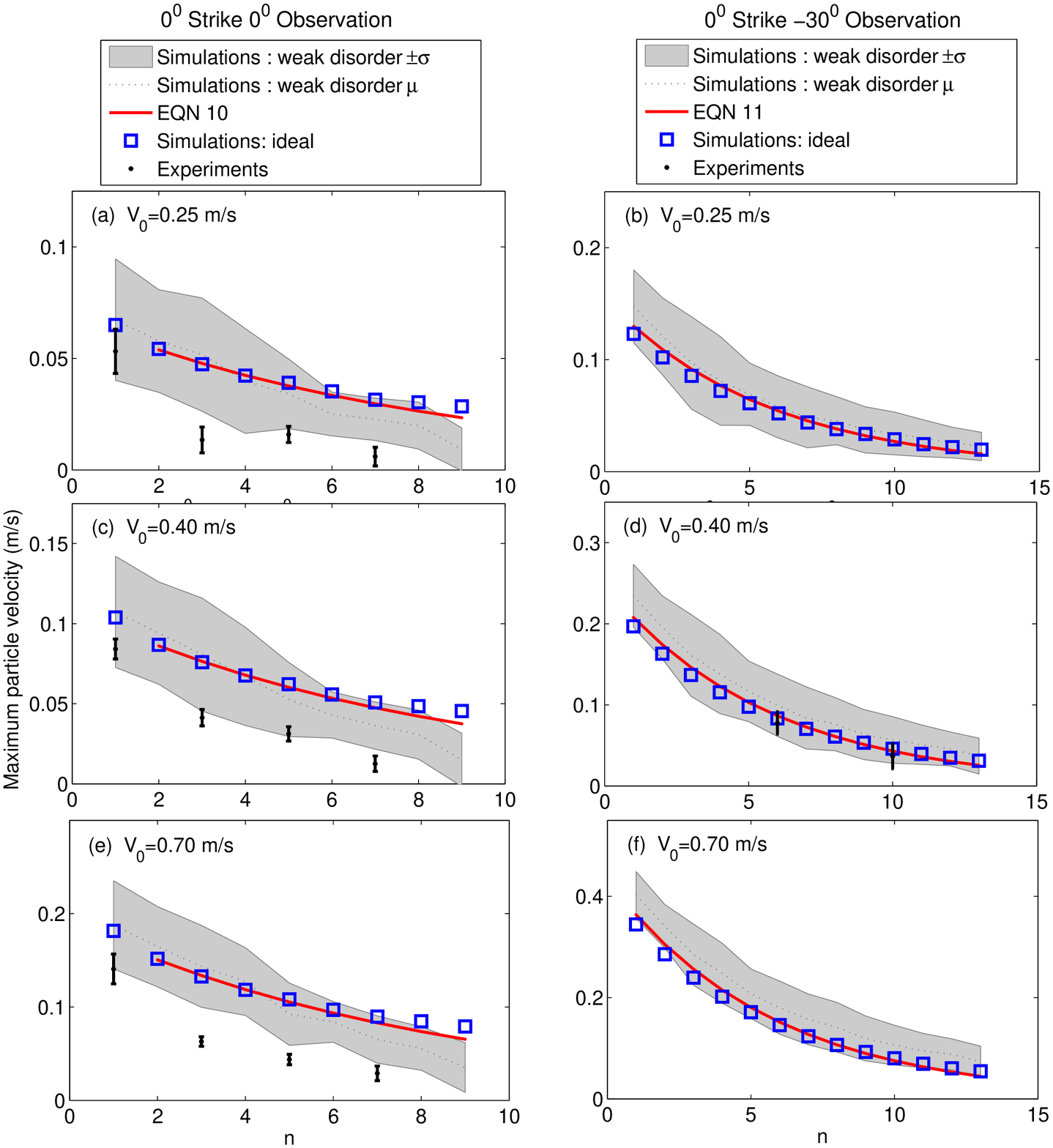, width=1.1 \textwidth} 
 }
 \caption{Maximum particle velocity with respect to sphere position denoted
by $n$. Panels (a), (c), (e): $0^0$ Strike and $0^0$ Observation. 
Panels (b), (d), (f): $0^0$ Strike and $-30^0$ Observation. Initial velocity: (a), (b) $V_{0}=0.25~\nicefrac[]{m}{s}$; (c), (d) $V_{0}=0.40~\nicefrac[]{m}{s}$; and (e), (f) $V_{0}=0.70~\nicefrac[]{m}{s}$. Shaded grey regions indicate  $\pm 1$ standard deviation, $\sigma$, and dotted grey lines indicate the average value, $\mu$, from the 20 simulations incorporating weak disorder. The solid red lines represent Equations \ref{eq:v0} and \ref{eq:veq0}. The black dots and error bars represent the mean and standard deviation from experiments ($V_{0}=0.25$, $0.40$, and $0.70~\nicefrac[]{m}{s}$ for $0^0$ Strike $0^0$ Observation and $V_{0}=0.40~\nicefrac[]{m}{s}$ for $0^0$ Strike $30^0$ Observation).}
 \label{fig:all_directions}
 \end{figure*}

%We presented several approaches to investigate the evolution of the leading propagating wave structure present in a 2D hexagonal granular crystal subjected to a localized impact. 
Based on the numerical simulations of the ideal hexagonal lattice (Figs. \ref{fig:smallsim} and \ref{fig:largesim}), the wavefront shape can be initially described by a hexagonal pattern which gradually 
transitions into a more circular shape.
This transition occurs as the scale of the structure changes from being
comparable to the lattice ``spacing'' to being much wider than that. It
is worthwhile to mention that this is an intriguing feature that perhaps
a quasi-continuum or homogenization type approach 
(see e.g.~\cite{craster} and references therein for recent work on the
subject) may capture and is left as an interesting open problem. Figure~\ref{fig:expnumsim} compares the initial wavefront shape observed in the ideal numerical simulations with those observed in the experiments and numerical simulations incorporating weak disorder. The average wave front arrival times from both experiments and numerical simulations incorporating weak disorder clearly show the initial hexagonal wave front propagating though the structure. The average arrival times in experiments are slightly longer than those predicted from the ideal numerical simulations. Physically, this can be understood since the dissipation present in experiments results in a decreased wave amplitude, consequently decreasing the wave front speed. Additionally, imperfections in the experimental contact lattices also result in redirection of wave amplitude and delays in the signal arrival time \cite{l11}. As Fig.~\ref{fig:expnumsim} indicates, the relative arrival times in the weakly disordered simulations are notably longer than both the predicted values from the ideal simulations as well as the experiments. This is a bit surprising considering that the simulations do not incorporate dissipation, thus the time delay is entirely due to the effects of the particle misalignments. The discrepancy between the arrival times in experiments and weakly disordered numerical simulations suggests that the tolerance values used to simulate imperfections in the contact lattice are fairly conservative.

Figure~\ref{fig:all_directions} compares the wave front amplitude decay along the $0^0$ and $-30^0$ observation directions after a $0^0$ strike angle for experiments, ideal numerical simulations, weakly disordered numerical simulations, and the corresponding Eqs. \eqref{eq:v0} and \eqref{eq:veq0}. Overall the trends observed for all approaches are in good agreement. The presence of weak disorder in the numerical simulations results in decreased amplitude transmission along the $0^0$ observation direction and increased amplitude transmission along the  $-30^0$ observation direction, compared to the ideal simulations. This altered distribution of wave front amplitude can be seen in Fig. \ref{fig:all_directions}, where the mean amplitude from simulations incorporating weak disorder falls above the ideal simulation values along the $\pm30^0$ observation direction and below the ideal simulation values along the $0^0$ observation direction.  The $30^0$ observation direction represents a line of spheres directly in contact, while the wave front must travel though a zig-zag of particle contacts along the $0^{0}$ direction. Therefore, the number of contacts, or possible amplitude scattering points, is greater along the $0^0$ observation direction, which could help physically explain this amplitude redistribution phenomenon. Although the presence of dissipation in experiments results in lower average amplitudes compared to the weakly disordered numerical simulations, which neglect dissipation, we also observe this trend in average experimental amplitude values (see the 
middle panel of Fig. \ref{fig:all_directions} for 
$V_{0}=0.40~\nicefrac[]{m}{s}$).

\section{Conclusions and Future Challenges}
\label{sec:theend}

We presented a systematic study of the dynamic response of a 2D hexagonal, highly
nonlinear lattice excited by an impulse. In this 2D hexagonal setting, because of the
ever-increasing number of neighbors over which the energy is
distributed, no genuine traveling wave excitations, i.e. with constant velocity, have been found
to persist. The propagating pulse energy has been found to decay as a power law, both in our numerical computations and in our experimental observations. Detailed expressions were provided to describe these power laws at
different angles of strike and of observation. 
In a special case (of $30^{\circ}$ strike
and $30^{\circ}$ observation, according to our presented classification),
a generalization of the binary collision approximation (dubbed the
ternary collision approximation) was presented and utilized to give
simple numerical and even approximate analytical expressions for the
bead evolution. Lastly, the effects of weak disorder on the propagating wave structure were examined; the average spatial amplitude values from numerical simulations incorporating weak disorder were in good agreement with experimental values as well as the corresponding fitted equations derived from numerical simulations on the ideal hexagonal granular crystal. This agreement reveals that the level of disorder present in experiments does not cause significant deviation of the propagating wave structure from the predicted system response.

While this is a fundamental step towards an improved understanding
of non-square lattices, there are numerous additional tasks to be
considered. While invariant traveling
solutions may not exist, the possibility of existence of 
self-similar decaying solutions of the discrete model or perhaps
of a quasi-continuum approximation thereof cannot be excluded
and should be further considered. Additionally, while here we
have concerned ourselves with the highly nonlinear limit of the model
(i.e., no precompression),
there has recently been a surge of activity in connection 
to linear or weakly nonlinear properties of non-square (such as
honeycomb or hexagonal) lattices. 
This is due to some of the remarkable linear properties of these lattices,
such as, e.g.,  conical diffraction and the existence of Dirac (diabolical)
points examined in both the physical~\cite{mark2,moti2} and mathematical~\cite{miw}
communities. It would be particularly relevant to consider the
possibility of such spectral and nonlinear features in the precompressed
variant of the present chain. Some of these possibilities 
are currently under investigation and will be reported in 
future publications.

\begin{acknowledgements}
This research is supported in part by the Department of Energy Office of Science Graduate Fellowship Program (DOE SCGF), made possible in part by  the American Recovery and Reinvestment Act of 2009, administered by ORISE-ORAU under contract no. DE-AC05-06OR23100, and the Army Research Office MURI grant US ARO W911NF-09-1-0436. P.G.K. acknowledges support from the US National 
Science Foundation under Grant No. CMMI-1000337, 
the US Air Force under Grant No. FA9550-12-1-0332, the 
Alexander von Humboldt Foundation, and the Alexander S.  Onassis 
Public Benefit Foundation.
\end{acknowledgements}

\end{document}